\newcommand{\ARAA}{ARA\&A}
\newcommand{\AaA}{A\&A}
\newcommand{\ApJ}{ApJ}
\newcommand{\MNRAS}{MNRAS}
\newcommand{\Natur}{Nature}
\newcommand{\PhLB}{Phys. Lett. B}
\newcommand{\PhL}{Phys. Lett.}
\newcommand{\PhRvD}{Phys. Rev. D}
\newcommand{\PhRvL}{Phys. Rev. Lett.}

\documentstyle[epsf]{aa}
\begin{document}
\def\gtsima{$\; \buildrel > \over \sim \;$}
\def\simgt{\lower.5ex\hbox{\gtsima}}
\thesaurus{12.03.1, 03.13.2, 03.13.6}
\title{Searching for the non-gaussian signature of the  
CMB secondary anisotropies}
\author{N. Aghanim \and O. Forni}
\offprints{N. Aghanim, aghanim@ias.fr}
\institute{IAS-CNRS, Universit\'e Paris Sud, B\^atiment 121, F-91405 Orsay 
Cedex}
\date{Received date / accepted date}
\maketitle
\markboth{Searching for the non-gaussian signature of the  
CMB secondary anisotropies}{}
\begin{abstract}
In a first paper (Forni \& Aghanim 1999), we developed several statistical 
discriminators
to test the non-gaussian nature of a signal. These tests are based on the 
study of the
coefficients in a wavelet decomposition basis. In this paper, we apply them 
in a cosmological context, to the study of the nature of the Cosmic Microwave 
Background (CMB) anisotropies. The latter represent the superposition of 
primary anisotropy imprints of the initial density perturbations and 
secondary ones due to photon interactions after recombination. In an 
inflationary scenario (standard Cold Dark Matter) with gaussian distributed 
fluctuations, we study the statistical signature of the
secondary effects. More specifically, we investigate the dominant effects 
arising from 
the Compton scattering of CMB photons in ionised regions of the Universe: the
Sunyaev-Zel'dovich effect of galaxy clusters and
the effects of a spatially inhomogeneous re-ionisation of the Universe.\par

Our study specifies and predicts the non-gaussian signature of the secondary 
anisotropies induced by these scattering effects. We find that our 
statistical discriminators allow us to distinguish and highlight the 
non-gaussian signature of a process even if it is combined with a larger 
gaussian one. We investigate the detectability of the secondary
anisotropy non-gaussian signature in the context of the future CMB 
satellites (MAP and Planck Surveyor). 
\end{abstract}
\section{Introduction}
The Cosmic Microwave Background (CMB) is a powerful tool for cosmology. 
As the CMB temperature anisotropies represent the superposition of 
primary (before matter-radiation decoupling) and secondary (after decoupling)
fluctuations, the study of the anisotropies gives a direct insight into both 
the early Universe (and initial conditions) and the formation and evolution of
cosmic structures. One of the goals of cosmology is to characterise 
the initial density perturbations which gave rise to those
structures: galaxies and galaxy clusters. The statistical properties of the
initial perturbations provide part of the necessary information for this 
characterisation. They can indeed be used to test and constrain the 
cosmological models and scenarios of structure formation.
The angular power spectrum of the temperature fluctuations is one of the most
important statistical quantities for CMB anisotropy studies. In
fact, it allows the evaluation of the main cosmological parameters
($\Omega_b$, $\Omega_0$, $\Lambda$, $n$, ...) defining our Universe 
(Jungman et al. 1996). Some of the first constraints on the cosmological
parameters came from CMB anisotropy measurements made by the
COBE satellite (Smoot et al. 1992; Wright et al. 1992). The statistical
properties of the CMB anisotropies give us information, in particular,
on the physical process at the origin of the initial density fluctuations. Two 
classes of
scenario account for the initial seeds of the structures. One is the
``inflationary model'' (Guth 1981; Linde 1982) in which the density 
perturbations result from the quantum
fluctuations of scalar fields in the very early Universe. The other 
invokes the topological defects which themselves correspond to symmetry
breaking in the unified theory (cosmic string, textures) 
(Vilenkin 1985; Bouchet 1988; Stebbins 1988; Turok 1989; Pen et al. 1994). 
Several studies 
have shown that the two scenarios predict different angular power spectra 
(Coulson et al. 1994; Albrecht et al. 1996; Magueijo et al. 1996). 
These differences of amplitude and/or
shape represent rather tight constraints on the models.
The statistical nature of the primary density perturbations, and hence their
origin, is also encompassed within the distribution of the CMB
anisotropies. The brightness, or temperature, distribution is indeed directly
induced by the primeval mass or density distribution. If 
the initial perturbations result from an inflationary process the primary 
anisotropy distribution is gaussian. If the perturbations are generated by 
topological defects the anisotropy 
distribution is non-gaussian. The latter predict very specific patterns 
distinguishable from a gaussian random field. 
It is thus necessary to find statistical methods to test non-gaussianity and 
to separate primary and secondary non-gaussianity.\par
Several studies have been 
performed to test the CMB gaussianity. Traditional methods use the brightness
or temperature distribution and their n$th$ order moments or their cumulants
(Ferreira et al. 1997). Other methods are based on 
the n-point correlation functions or their spherical harmonic transforms 
(Luo \& Schramm 1993; Magueijo 1995; Kogut et al. 1996; Ferreira \& Magueijo 
1997; Ferreira et al. 1998; Heavens 1998; Spergel \& Goldberg 1998). 
Non-gaussianity can also be tested
through topological discriminators based on pattern statistics 
(Coles 1988; Gott et al. 1990). Alternative methods test the non-gaussianity 
in the Fourier or wavelet space (Ferreira \& Magueijo 1997; Hobson et al. 
1998; Forni \& Aghanim 1999). 
\par
In addition to the intrinsic statistical properties of the CMB anisotropies,
the secondary fluctuations associated with cosmic structures (e.g., galaxies
and galaxy clusters) induce non-gaussian signatures which could originate from 
point-like sources, peaked profiles,
or from geometrical characteristics such as sharp 
edges or specific patterns. Future high sensitivity and high resolution
CMB observations (e.g., MAP\footnote{http://map.gsfc.nasa.gov/} and Planck 
Surveyor\footnote{http://astro.estec.esa.nl/SA-general/Projects/Planck/} 
satellites) will provide data sets which should allow detailed tests of 
the primary anisotropy distribution. A detailed
study of the non-gaussianity associated with secondary sources could be used
to discriminate between the inflationary and topological defect
models. \par
The present study deals with this first step: to predict and to specify 
the non-gaussian signature of the secondary anisotropies arising
from the scattering of CMB photons by the ionised matter in the Universe.
We apply the statistical discriminators developed in
Forni \& Aghanim (1999) to combinations of gaussian primary and 
secondary non-gaussian anisotropies. We take into account the contribution of a
population of galaxy clusters through the Sunyaev-Zel'dovich (SZ) effect 
(Sunyaev \& Zel'dovich 1980) as well as the effect of a spatially 
inhomogeneous 
re-ionisation of the Universe (Aghanim et al. 1996; Gruzinov \& Hu 1998;
Knox et al. 1998). The non-gaussian signature 
due to secondary anisotropies associated with weak gravitational lensing
have been investigated in previous studies 
(Seljak 1996b; Bernardeau 1998; Winitzki 1998).
\par
In section 2, we present the astrophysical contributions we take into account 
 in our study. We then briefly present the statistical tests and detection
strategy in section 3. We apply our tests to the combinations of
primary and secondary anisotropies due to inhomogeneous re-ionisation alone
in section 4, and to a configuration including the SZ effect of galaxy
clusters in section 5.
In section 6, we investigate the detectability of the non-gaussian signature 
for a MAP-like and a Planck-like
instrumental configuration. Finally, in section 7, we discuss our results and 
present our conclusions.
\section{Astrophysical contributions}
The temperature anisotropies of the CMB contain the contributions 
of both the primary cosmological signal, directly related to the initial 
density fluctuations, and
the foreground contributions amongst which are the secondary anisotropies. 
The secondary anisotropies are generated after matter-radiation decoupling. 
They arise
from the interaction of the CMB photons with the matter and
can be of a gravitational type (e.g. Rees-Sciama effect (Rees \& Sciama 1968)),
 or of a scattering type when the matter is ionised (e.g. SZ or
Ostriker-Vishniac effect (Ostriker \& Vishniac 1986; Vishniac 1987)). In our 
study we adopt a canonical inflationary standard
CDM (Cold Dark Matter) model for the generation of the primary anisotropies. 
\par
We simulate maps of the three astrophysical processes 
of interest in our study: the primary and secondary fluctuations due
to inhomogeneous re-ionisation and the SZ effect. For each process, we made 
100 realisations of $512\times512$ pixels 
(1.5 arcminute pixel size). This fairly large number of realisations allows 
us to have statistically significant results. They represent about 40\% of the 
whole sky, 
equivalent to the ``clean'' fraction of the sky coverage available for CMB 
analysis.
Indeed, we do not expect to be able to analyse regions of the sky that 
are highly contaminated by galactic emissions (dust, synchrotron and free 
free). These contaminated regions account, more or less, for the Galactic 
latitudes with 
$|b|<30^{\circ}$ ($\sim60$\% of the sky).
\subsection{Primary CMB anisotropies}
For the purpose of this study, that is the characterisation of the 
non-gaussianity from secondary 
anisotropies, we assume gaussian distributed primary fluctuations generated in 
an inflationary scenario. We choose the canonical
standard CDM model, normalised to COBE data. The maps were generated using 
a code, kindly provided by P.G. Ferreira that generates square gaussian 
realisations given a power spectrum. The CMB power spectrum, displayed in 
figure \ref{fig:cl}, was computed using
the CMBFAST code (Seljak \& Zaldarriaga 1996). \\
\subsection{Secondary CMB anisotropies}
\subsubsection{From inhomogeneous re-ionisation}
The first generation of emitting objects
ionises the surrounding gas of the globally neutral Universe at high
redshifts. The resulting spatially inhomogeneous re-ionisation generates 
secondary anisotropies
associated with the peculiar motion along the line of sight of ionised
bubbles. This produces anisotropies with maximum amplitude at the
degree scale, and with $(\delta T/T)_{rms}\sim 6.\,10^{-6}$. The anisotropies 
are about ten 
times smaller than the primary fluctuations and spectrally indistinguishable
from them. We use the model of Aghanim et al. (1996), in which these
objects are early ionising quasars with assumed lifetimes of $10^7$ yrs. 
The number 
of quasars is normalised to match the data at $z\sim4$ and has been 
extrapolated for $4<z\le10$. The
positions of the centres of the ionised regions are drawn at random in the 
$512\times512$
pixel maps, and we assume a spherically symmetric gaussian profile for the 
temperature anisotropy. The
size and amplitude of the anisotropies depend on the quasar luminosities
and its light-on redshifts. We compute the
skewness and kurtosis (third and fourth moments of the distribution) of the 
maps. All the maps exhibit a non-gaussian signature associated with an excess 
of kurtosis of the order of one, the skewness being null.
\subsubsection{From Sunyaev-Zel'dovich effect} 
The Sunyaev-Zel'dovich effect 
represents the Compton scattering of the CMB photons by the free electrons of
the ionised and hot intra-cluster gas. It results in the so-called thermal SZ
effect which exhibits a peculiar 
spectral signature with a minimum at long wavelengths and a 
maximum at short wavelengths. When the cluster moves with respect to the CMB 
rest 
frame, the Doppler shift induces an additional effect called the kinetic SZ
effect. It generates anisotropies with the same spectral signature as the 
primary ones.
The temperature anisotropies generated by the clusters are thus 
composed of the thermal $(\delta T/T)_{th}$ and kinetic $(\delta T/T)_{ki}$ 
SZ anisotropies. We simulate both
effects using an updated version (Aghanim et al. 1998) of the Aghanim et al 
(1997) model. The 
simulations use the $\beta$-model (Cavaliere \& Fusco-Femiano 1978) to describe the 
gas distribution of each individual cluster. This description is
generalised to a population of clusters derived from the Press-Schechter 
formalism (Press \& Schechter 1974) and normalised to the X-ray temperature 
distribution (Viana \& Liddle 1996). The positions of the cluster centres are
drawn at random in the maps. Again we find a zero skewness 
but a strongly non-gaussian signature because the kurtosis is non-zero.
\section{Statistical tests and detection strategy}\label{sec:met}
In a previous study (Forni \& Aghanim 1999), we developed 
statistical methods to search for non-gaussianity. The tests are based on the
detection of gradients in the wavelet space. They use
the statistical properties of a signal in wavelet space, namely
the measurement of the excess of kurtosis (fourth moment, $\mu_4$, of a 
distribution) of the coefficients associated with the gradients. The predicted 
excess of kurtosis for a gaussian distribution is zero. If the non-gaussian 
signal is not skewed (third moment of the distribution is zero), any 
significant departure from gaussianity is indicated by a non-zero excess of 
kurtosis.
\begin{figure}
\epsfxsize=\columnwidth
\hbox{\epsffile{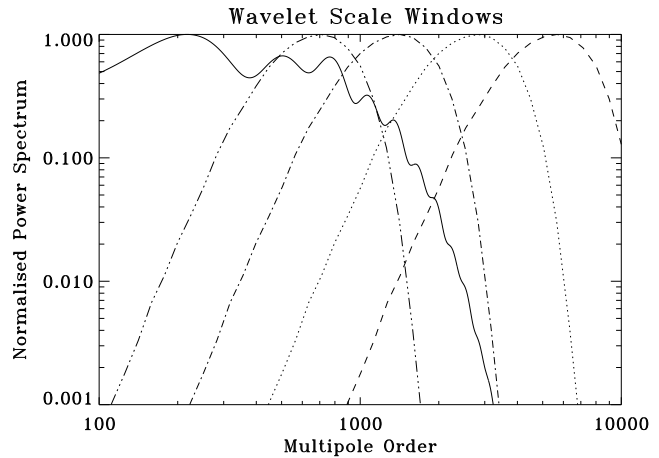}}
\caption{{\small\it  Scale windows corresponding to the 4 decomposition levels 
(in the spherical harmonics space) by the bi-orthogonal filter \#3 
(Villasenor et al. 1995). From right (scale I) to left (scale IV), the windows 
are centred around a multipole of 
$\sim5600$,$\sim2800$, $\sim1400$ and $\sim700$. 
Their respective Full Widths at Half Maximum are $(\Delta l)_{FWHM}\sim4800$, 
$\sim2400$, $\sim1200$ and $\sim600$. The solid 
line shows the normalised CMB primary anisotropy power spectrum.}}
\label{fig:cl}
\end{figure}
\par
The first test for non-gaussianity is based on what we call the multi-scale
gradient. It is the quadratic sum of the wavelet coefficients associated with  
$(\partial/\partial x)^2+(\partial/\partial y)^2$ and we find it follows a
Laplace distribution for a gaussian distributed signal. For sets of 100 
statistical realisations wavelet
filtered at four decomposition scales (Fig. \ref{fig:cl}), we compute the 
normalised excess of kurtosis with respect to the Laplace distribution 
($k=\mu_4/\mu_2^2-6$) together with the standard deviations (with respect to 
the median excess of kurtosis). 
\par
The second statistical test uses the wavelet
coefficients associated with the horizontal and vertical gradients 
($\partial/\partial x$ and $\partial/\partial y$ derivatives). We also use the
coefficients related to the diagonal gradients 
($\partial^2/\partial x\partial y$ 
cross derivative), which are gaussian distributed for a gaussian signal. We
then compute the normalised
excess of kurtosis ($k=\mu_4/\mu_2^2-3$), with respect to a gaussian 
distribution and the standard deviations (with respect to the median) for all
the realisations. 
\par
We apply the detection strategy proposed in Forni \& Aghanim (1999) to demonstrate 
and to quantify the detectability of the non-gaussian signature. It is based 
on the comparison of a set of maps of the ``real'' observed sky to a set
of gaussian realisations having the power spectrum of the ``real sky''.
The main 
advantage of this is that it can be reliably applied regardless of the power
spectrum of the studied non-gaussian signal.
\section{Analysis of the anisotropies: Primary + inhomogeneous re-ionisation}
We first study the case of primary CMB anisotropies with secondary 
anisotropies due to inhomogeneous re-ionisation. The 
primary CMB anisotropies dominate at all scales larger than the cut off (at 
about
5 arcminutes). The non-gaussian signal is very small compared to the gaussian
one. Indeed, the power spectrum of the secondary anisotropies 
represents, at most, less than 10\% of the primary CMB power. 
\subsection{Multi-scale gradient}
We compute the median value of the 
excess of kurtosis for the 100 realisations and the 
associated confidence intervals (Tab. \ref{tab:cmb_bng}). At the first 
decomposition scale the
secondary anisotropies dominate the primary. We thus
expect the non-gaussian signature of the secondary anisotropies to dominate,
and indeed, we find a non-zero excess of kurtosis.
At the two larger decomposition scales, the median 
value $k$ is marginally non-zero. The computed $\sigma$ values 
take into account the non symmetrical distribution of the kurtosis and 
exhibit a clear dichotomy between the upper ($\sigma_+$) and lower ($\sigma_-$)
boundaries of the confidence interval. This suggests that non-gaussianity has
been detected. If 
$k-\sigma_-$ for one realisation is larger than zero by a value of the order 
of, or larger than,
$\sigma_-$; then this indicates a significant departure 
from gaussianity. If $k-\sigma_-$ is of the order of zero then, more 
sophisticated tests must be applied to conclude weather the ``real sky'' has
a non-gaussian signature. At the second decomposition scale, the non-zero 
value of the median excess of kurtosis is possibly due to the sampling effects 
resulting from the sharp cut off in the primary CMB power spectrum at about 5 
arcminutes (in a standard CDM model) combined with the rather narrow window 
filter we use in the analysis.
\begin{table}
\begin{center}
\begin{tabular}{|c|c|c|c|}
\hline
Scale 	& $k$	& $\sigma_+$ & $\sigma_-$   \\
\hline
I     	& 0.3	& 0.15		& 0.16 \\
II    	& 1.19	& 0.88 		& 0.50  \\
III   	& 0.15  & 0.67	 	& 0.41   \\
\hline
\end{tabular}
\end{center}
\caption{\small\it The median excess of kurtosis $k$ of the multi-scale 
gradient coefficients, at four decomposition scales, computed over 100 
realisations of the CMB primary + secondary anisotropies due to inhomogeneous 
re-ionisation. $\sigma_+$ $\sigma_-$ and define the confidence interval for 
one realisation.}
\label{tab:cmb_bng}
\end{table}
\subsection{Partial derivatives} \label{sec:derv_re}
\begin{table}
\begin{center}
\begin{tabular}{|c|c|c|c|c|c|c|c|}
\hline
Scale & & $k_1$ & $\sigma_+$ & $\sigma_-$ & & $k_2$ & $\sigma_{\pm}$ \\
\hline
I & $\partial/\partial x$ & 0.04 & 0.02 & 0.02 & & -$10^{-4}$ & 0.02 \\
II & \& & 0.12 & 0.06 & 0.05 & $\partial^2/\partial x\partial y$ & 0.06 & 0.04 \\
III & $\partial/\partial y$ & 0.02 & 0.09 & 0.07 & & 0.02 & 0.08 \\
\hline
\end{tabular}
\end{center}
\caption{\small\it The median excess of kurtosis, at four decomposition 
scales, computed over 100 realisations of the sum of CMB and
secondary anisotropies due to inhomogeneous re-ionisation. $k_1$ is given for
the coefficients
associated with the vertical and horizontal gradients, and $k_2$ is given for 
the cross derivative. The $\sigma$ values
are the boundaries of the confidence interval for one statistical realisation.}
\label{tab:cbn_derv}
\end{table}
The analysis of the coefficients, associated with the cross 
and the first derivatives (Tab. \ref{tab:cbn_derv}), exhibit non-zero median 
excess of kurtosis at the 
first two decomposition scales. At the third and fourth scales the obtained 
excess of kurtosis is very close to the values of the CMB primary 
anisotropies alone. 
\begin{figure*}
\hbox{\epsffile{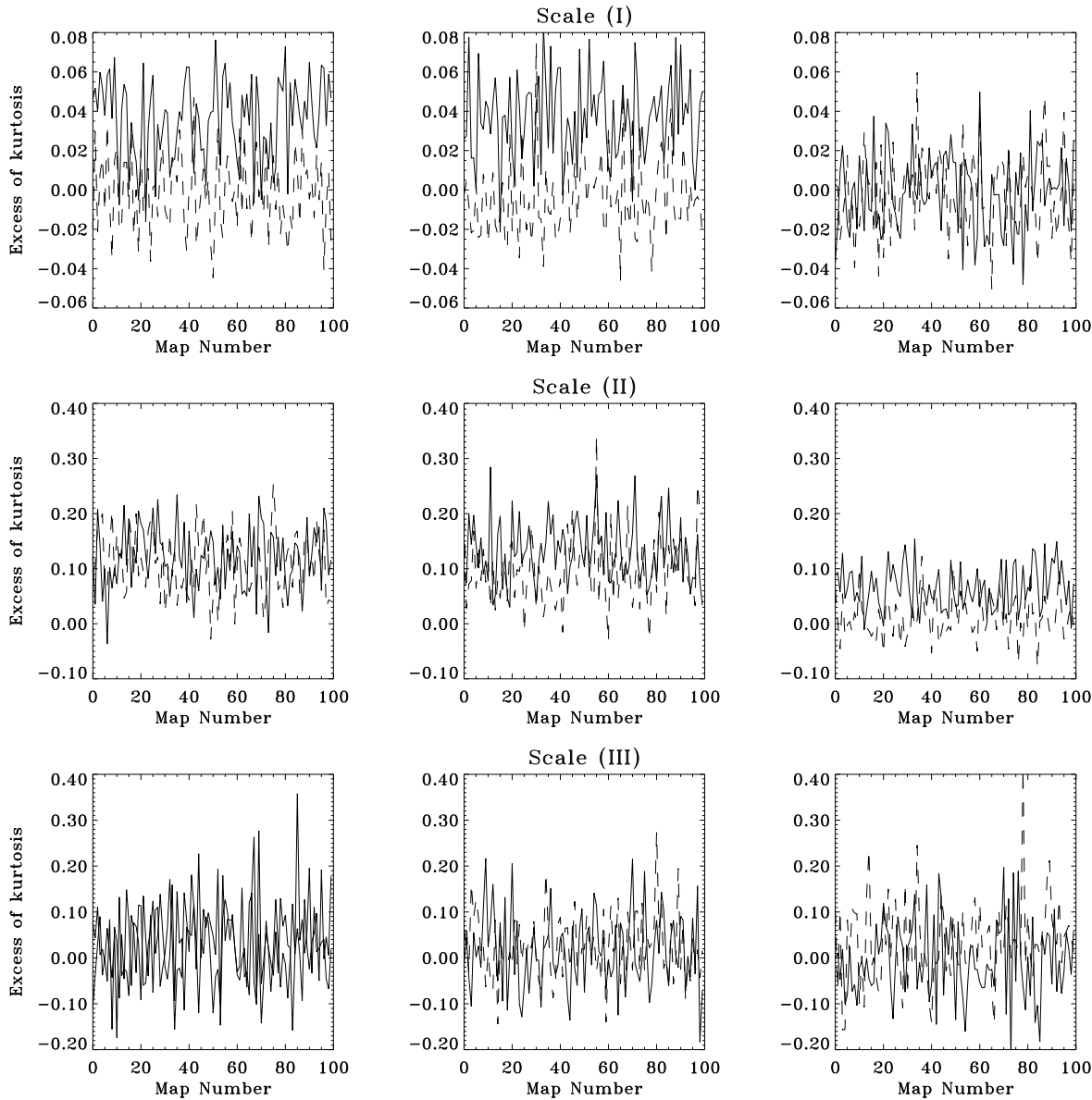}}
\caption{{\small\it Excess of kurtosis computed over the wavelet coefficients
of respectively $\partial/\partial x$ for the left panels, 
$\partial/\partial y$ for the centre panels and $\partial^2/\partial x
\partial y$ for the right panels. The solid line is for CMB + non-gaussian
secondary anisotropies due to patchy re-ionisation. The dashed line is for 
gaussian realisations, with the same power spectrum.}}
\label{fig:mom_cbgn}
\end{figure*}
In order to illustrate the non-gaussian characteristics of the different
statistical realisations, 
we plot (Fig. \ref{fig:mom_cbgn}) the excess of kurtosis of the wavelet
coefficients associated with the partial derivatives 
($\partial/\partial x$ and $\partial/\partial y$) and the cross derivative
$\partial^2/\partial x\partial y$. The solid
line corresponds to the results obtained for the
100 realisations of a non-gaussian process made of the primary CMB + secondary 
anisotropies. Whereas the dotted line stands for the 
100 gaussian test maps with same power spectrum as the studied process. The 
excesses of
kurtosis are computed with the coefficients related to the horizontal gradient
$\partial/\partial x$ (left panels), 
to the vertical gradient $\partial/\partial y$ (centre panels), and to the 
cross derivative $\partial^2/\partial x
\partial y$ (right panels). For
the gaussian signal, the excesses of kurtosis of the cross derivative
coefficients are centred around zero whereas they are not for the first 
derivative coefficients. This indicates that the cross derivative coefficients
better characterise gaussian signals. They thus seem more appropriate to test
for non-gaussianity, even though the coefficients have 
smaller amplitudes. The results show a clear departure of the excess 
of kurtosis from
zero at the two first decomposition scales, for the non-gaussian signal. 
At the third decomposition
scale, the excess of kurtosis for the cross derivative becomes very weak,
indicating  a marginal detection of non-gaussianity. 
\par
\begin{figure}
\epsfxsize=\columnwidth
\hbox{\epsffile{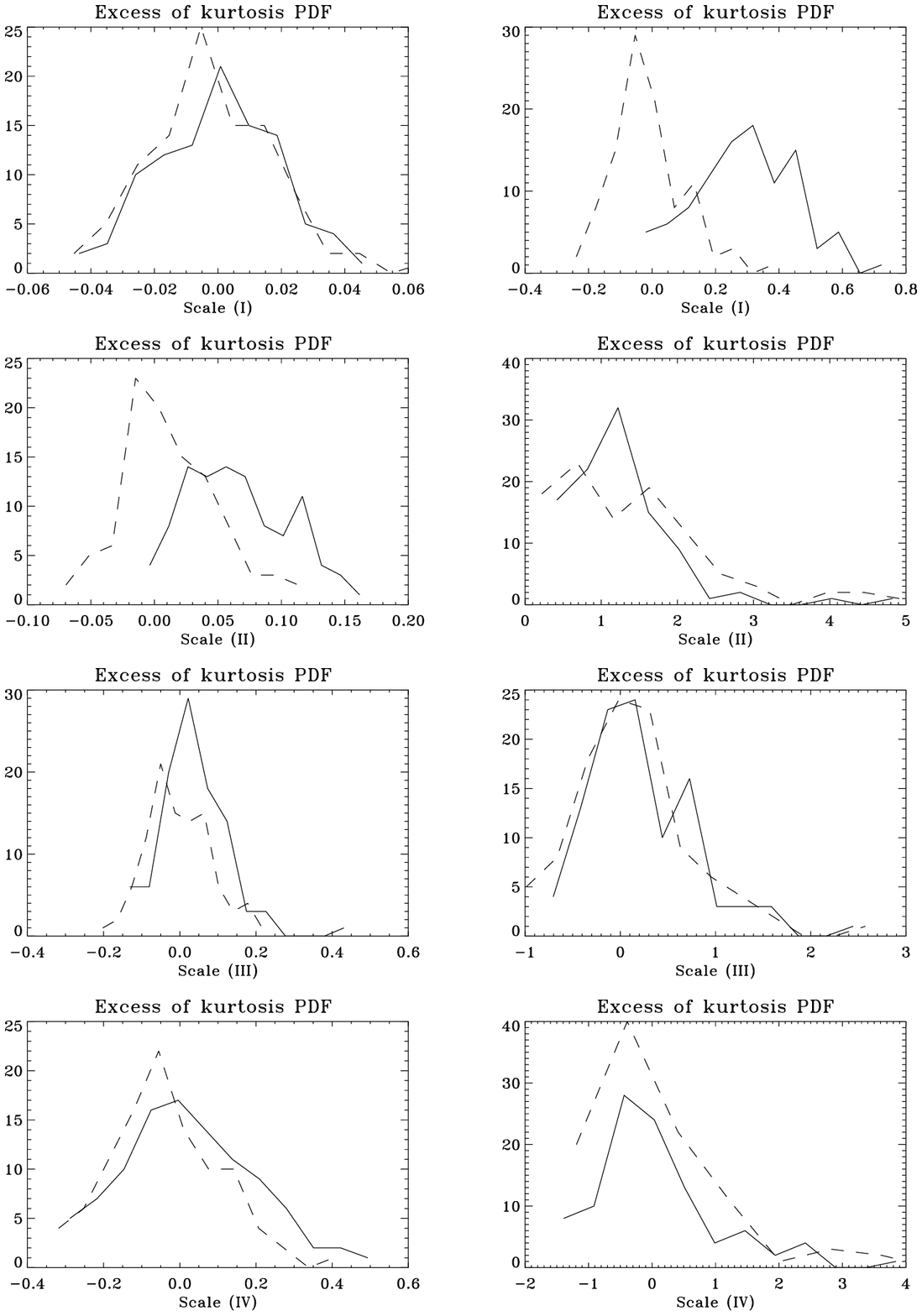}}
\caption{{\small\it Probability distribution functions, in percentage, of the 
excess of 
kurtosis computed with the multi-scale gradient coefficients (right panels)
and with the cross derivative coefficients (left panels). The dashed line is
for the gaussian test maps with same power spectrum as the non-gaussian signal 
(solid line). This signal is made of primary CMB anisotropies and secondary
anisotropies due to inhomogeneous re-ionisation.}}
\label{fig:pdfcb}
\end{figure}

Following the detection strategy of Forni \& Aghanim (1999), we perform 
a set of 100 gaussian realisations with same power spectrum as the sum of
CMB and inhomogeneous re-ionisation, and we compute the excess of 
kurtosis for
both the multi-scale gradient and the partial derivative coefficients. The 
probability distribution function (PDF) of the excess of kurtosis, for the 
non-gaussian (solid line) and gaussian (dashed line) realisations, are plotted
in figure \ref{fig:pdfcb}. 
We display, in the left panels, the PDF for the cross derivative 
coefficients. In the right panels, we show the excess of kurtosis computed with
the multi scale gradient coefficients. We note that investigating the 
statistical properties with the multi-scale gradient coefficients and 
cross derivative coefficients is quite complementary,
because the multi scale gradient is related to $\partial/\partial x$ and
 $\partial/\partial y$. The detection of the non-gaussian
signature is clear when the PDFs are clearly 
shifted. Our results show that the median excess of kurtosis of 
the multi scale gradient coefficients measures the non-gaussian nature of the
secondary anisotropies due to inhomogeneous re-ionisation with a probability 
of 99.76\% at the first decomposition scale. At all other scales, the
detection level, for this discriminator, is below the one sigma  
limit. For the statistical test based on the cross derivative, the probability
that the non-gaussianity is detected is 89.5\% at the second decomposition 
scale. All the other scales show no significant departure from gaussianity.
\section{Analysis of the anisotropies: including the SZ effect}
Besides the secondary anisotropies that would arise in the context of an 
inhomogeneous re-ionisation of the Universe, there exist secondary 
anisotropies due to the SZ effect of galaxy 
clusters. In our study, we therefore add to the previous model 
the contribution of both thermal and kinetic SZ effects of a
galaxy cluster population. We analyse the maps corresponding to the sum 
of CMB primary 
and secondary fluctuations (SZ + inhomogeneous re-ionisation) with a resolution
of 1.5 arcminutes and the nominal Planck gaussian noise. The contribution of
the thermal SZ effect, $(\delta T/T)_{th}$, is given at 2mm.  
\subsection{Multi-scale gradient}
We compute the excess of kurtosis of the multi-scale gradient coefficients of 
the primary + secondary anisotropy maps. In Table 
\ref{tab:cmb_bsz1.5}, we note the extraordinarily large values of the median
excesses of kurtosis $k$ with respect to the previously studied process
(CMB primary anisotropies + secondary fluctuations due to inhomogeneous
re-ionisation). More specifically, 
at the first and second decomposition scales the excess of kurtosis is 
respectively of the 
order 2300 and 180. At the third scale, we find $k=0.87$ which is already 
almost eight times greater than the corresponding value in Tab. 
\ref{tab:cmb_bng}. At the fourth
and largest scale, the excess of kurtosis is very small; it is comparable to 
that measured without the SZ anisotropies, very close to the CMB alone.
The non-gaussian signature, exhibited by the excess of kurtosis of the 
multi-scale gradient, is thus dominated at the first three scales by the SZ 
effect contribution, even though the latter does not 
dominate in terms of power. 
\begin{table}
\begin{center}
\begin{tabular}{|c|c|c|c|}
\hline
 Scale & $k$ & $\sigma_+$ & $\sigma_-$\\
\hline
  I & 2299.13 & 3793.56 & 1121.44\\
 II & 178.92 & 586.02 & 80.55 \\
 III & 0.87 & 1.39 & 0.63 \\
\hline
\end{tabular}
\end{center}
\caption{\small\it The median excess of kurtosis $k$, at four decomposition 
scales, computed over the 100 combinations of the sum of CMB and 
secondary anisotropies. The $\sigma$ values are the $rms$ values 
with respect to the median excess of kurtosis for one realisation.}
\label{tab:cmb_bsz1.5}
\end{table}
\subsection{Partial derivatives}
\begin{figure*}
\hbox{\epsffile{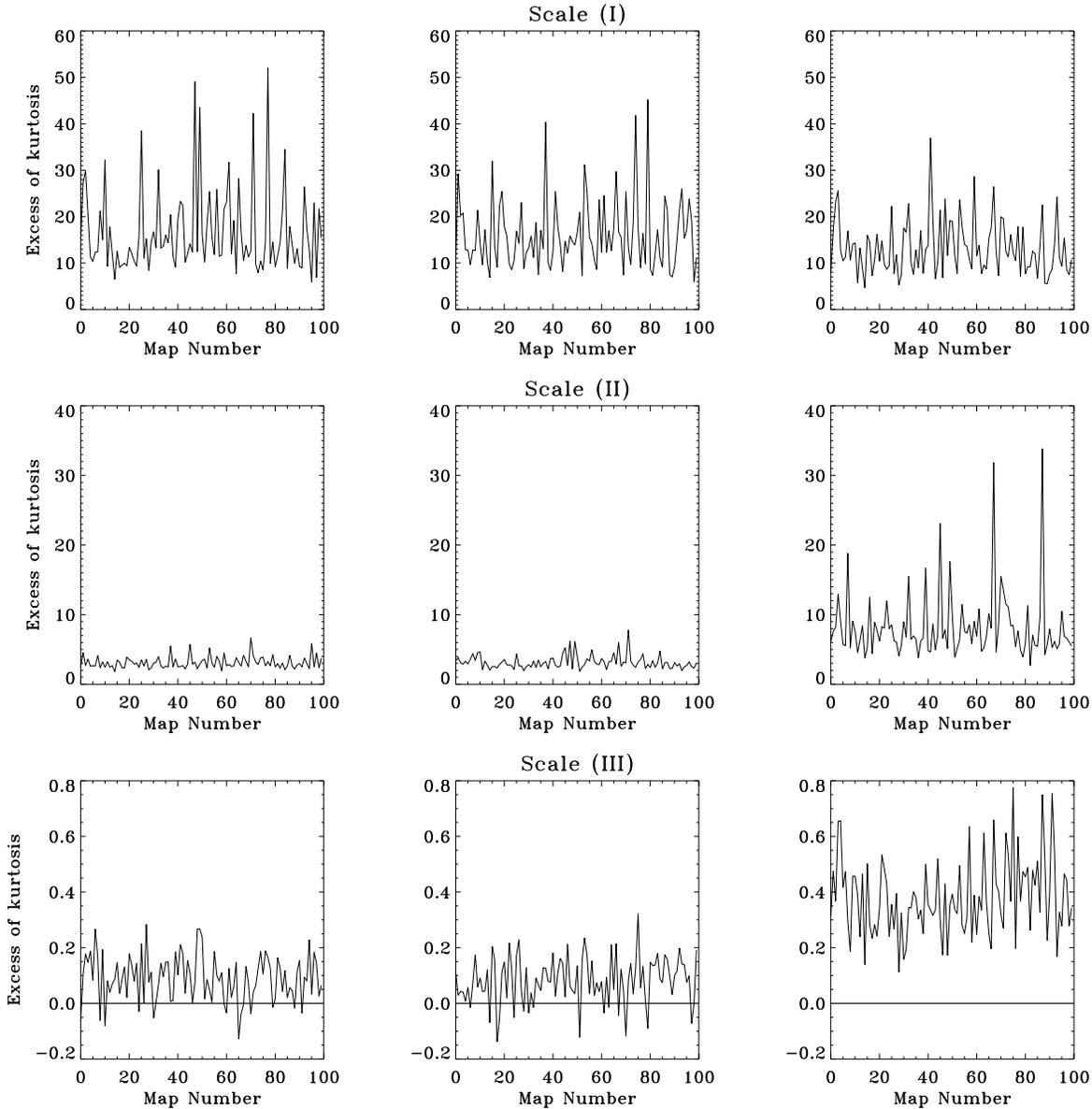}}
\caption{{\small\it Excess of kurtosis computed over the wavelet coefficients
of, respectively, $\partial/\partial x$ for the left panels, 
$\partial/\partial y$ for the centre panels and $\partial^2/\partial x
\partial y$ for the right panels. The signal includes the CMB + non-gaussian
secondary anisotropies due to patchy re-ionisation and the SZ effect. The
signal also includes gaussian noise at the level expected for the Planck 
mission.}}
\label{fig:mom_cbsz15}
\end{figure*}
For the same test maps, we compute the excess of kurtosis using the wavelet 
coefficients 
associated with the first and cross partial derivatives of the signal 
(Tab. \ref{tab:cbsz_derv} and Fig. 
\ref{fig:mom_cbsz15}). At the first three decomposition
scales, the excess of kurtosis is very large due to the SZ 
contribution. We also note, in agreement with our suggestions of Sec.
\ref{sec:derv_re}, 
that the computations using the cross partial derivative are more sensitive 
to non-gaussianity and thus more powerful in detecting it. In 
fact, the galaxy clusters exhibit very peaked profiles or even point-like 
behaviour. The wavelet coefficients associated with
the cross derivative, which are very sensitive to symmetric
profiles, are thus larger than in the previous study (inhomogeneous 
re-ionisation alone) in which we assumed a gaussian profile.
\begin{table*}
\begin{center}
\begin{tabular}{|c|c|c|c|c|c|c|c|c|}
\hline
Scale & & $k_1$ & $\sigma_+$ & $\sigma_-$ & & $k_2$ & $\sigma_+$ & $\sigma_-$\\
\hline
I & $\partial/\partial x$ & 13.89 & 11.67 & 3.97 & & 11.98 & 7.66 & 3.74 \\
II &  \& & 2.90 & 1.29 & 0.48 & $\partial^2/\partial x\partial y$ & 6.88 & 7.00 & 1.73\\
III & $\partial/\partial y$ & 0.09 & 0.08 & 0.08 & & 0.36 & 0.17& 0.11 \\
\hline
\end{tabular}
\end{center}
\caption{\small\it The median excess of kurtosis, at four decomposition 
scales, computed over 100 realisations of the sum of CMB and
secondary anisotropies (inhomogeneous re-ionisation and thermal and kinetic SZ 
effect). The signal includes the gaussian noise expected for the Planck 
mission. $k_1$ is the median excess computed with the coefficients
associated with the vertical and horizontal gradients, and $k_2$ is given for 
the cross derivative. The $\sigma$ values are the boundaries
of the confidence interval for one statistical realisation.}
\label{tab:cbsz_derv}
\end{table*}
\par
\begin{figure}
\epsfxsize=\columnwidth
\hbox{\epsffile{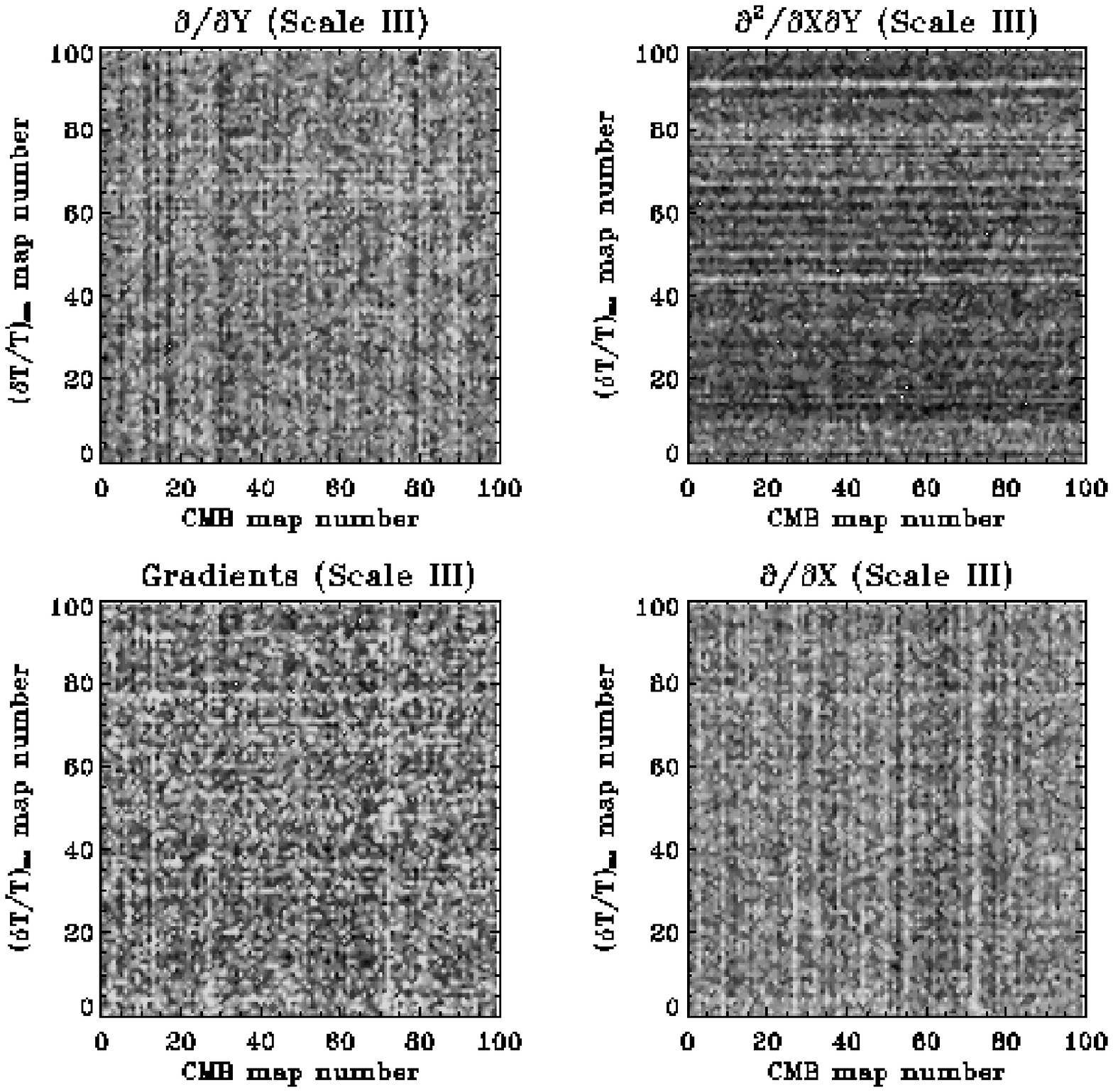}}
\caption{{\small\it At the \underline{third decomposition scale}, excess 
of kurtosis of $\partial/\partial x$, $\partial/\partial y$ and $\partial^2/
\partial x\partial y$ coefficients, and multi-scale gradient coefficient. The 
$x$ and $y$ axes represent the number of maps. The studied secondary 
anisotropies are those due to both the inhomogeneous re-ionisation and the
thermal and kinetic SZ effect + noise. The horizontal lines indicate the 
dominating non-gaussianity of the secondary anisotropies. The vertical lines
exhibit the dominating primary CMB signal.}}
\label{fig:cad3_cbsz}
\end{figure}

\bigskip
We illustrate, in Figure \ref{fig:cad3_cbsz}, the departure of the 
excess of kurtosis from zero. The $x$ and $y$ axes represent respectively, 
the number of the secondary and of the CMB primary anisotropy maps.  
The upper left and lower right images represent the excess of
kurtosis computed with the coefficients of 
$\partial/\partial x$ and $\partial/\partial y$. The upper right images were
obtained with the coefficients of $\partial^2/\partial x\partial y$. The lower 
left image shows the excess of kurtosis computed with the multi-scale 
gradient coefficients. Up to the third scale, the horizontal 
lines dominate the image, outlining a highly non-gaussian signal due to 
the secondary anisotropies. This is particularly true for the cross
derivative coefficients (top right image). The other statistical
tests show the non-gaussian secondary anisotropies but they also exhibit the
CMB associated features (vertical lines). 
\section{Effects of the instrumental configurations}
We apply our statistical discriminators to test for non-gaussianity within the 
context of the representative instrumental configurations of the future 
MAP and Planck Surveyor satellite for CMB observations. The Planck 
configuration allows an investigation of the beam convolution effects alone, 
because
the noise level remains unchanged. Whereas for the MAP configuration, we vary
both the beam and the noise level.
\subsection{MAP-like configuration}
We have used a MAP-like instrumental configuration corresponding to a
convolution, with a gaussian beam of full width at half maximum of 12 
arcminutes, of maps consisting of the primary CMB anisotropies to which we
added the secondary anisotropies. 
The noise added to the convolved maps is gaussian with $rms$ amplitude
$(\delta T/T)_{rms}=10^{-5}$. From these ``observed'' maps, we compute the 
median excess of kurtosis for the multi-scale gradient coefficients and for the
coefficients of the first and cross derivatives.
At the first two decomposition scales the signal is suppressed due to the beam 
dilution effects and the fourth decomposition
scale is dominated by the CMB primary anisotropies. In the
MAP-like configuration, we are therefore left with a unique decomposition
scale, the third, to test non-gaussianity using our methods.
At this scale, the excess of kurtosis for the cross derivatives is 
$0.07\pm0.03$. Whereas it is rather large for the multi-scale gradient, and 
first derivatives respectively, 
$1.15^{+2.24}_{-0.78}$ and $0.14^{+0.13}_{-0.10}$. Here again, the 
non-zero excess of kurtosis is possibly due to a sample variance problem, as 
the 12 arcminute convolution sharply cuts the power at the 
third decomposition level (Fig. \ref{fig:cl}). Using the PDF of the excess of 
kurtosis, we compute the probability that
the measured excess belongs to a non-gaussian signal and we find it below the
one sigma detection limit. We also apply the Kolmogorov-Smirnov (K-S) test 
(Press et al. 1992)
which compares globally two distribution functions, especially the shift in the 
median value. We find that the PDFs of the gaussian process and of the ``real
sky'' observed by MAP are identical. These results, using our statistical 
discriminators, 
thus suggest that the MAP satellite will be unable to detect non-gaussianity.
\subsection{Planck Surveyor-like configuration}
\begin{figure}
\epsfxsize=\columnwidth
\hbox{\epsffile{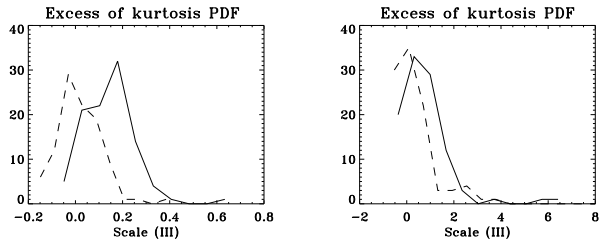}}
\caption{{\small\it For the Planck-like configuration (CMB + SZ + inhomogeneous
 re-ionisation), probability distribution functions, as percentages, of the 
excess of 
kurtosis computed with the multi-scale gradient coefficients (right panels)
and with the cross derivative coefficients (left panels). The dashed line is
for the gaussian test maps with same power spectrum as the non-gaussian signal 
(solid line).}}
\label{fig:pdfpl}
\end{figure}

We use the same astrophysical contributions as those of the 
MAP-like configuration (primary and SZ + inhomogeneous re-ionisation secondary 
anisotropies). These maps are convolved 
with a 6 arcminutes gaussian beam. We also take into account the expected 
gaussian noise of Planck ($(\delta T/T)_{rms}\sim2.\,10^{-6}$ per 1.5 
arcminute pixel). The convolution by a 6 arcminute beam suppresses the 
power at the corresponding scale (Scale I) and affects the second 
decomposition scale. The third one is not significantly 
altered by the convolution and we expect that the non-gaussianity could thus 
be detected. For the multi-scale gradient we find $k=0.62^{+1.43}_{-0.60}$.
Whereas we find for the first and cross derivatives respectively, 
$k=0.07^{+0.11}_{-0.08}$ and $0.16\pm0.10$.  In order to quantify the 
detectability of the non-gaussianity
in the Planck-like configuration, we generate gaussian distributed maps with 
same power spectrum as the 
studied signal. We plot (Fig. \ref{fig:pdfpl}) the PDF of the gaussian (dashed 
line) and non-gaussian (solid line) 
processes. We derive the probability that the median excess of kurtosis 
measured on the ``real sky'' belongs to
the gaussian process. Using the multi-scale gradient we find that the 
probability of detecting non-gaussianity is
71.9\% at the second decomposition scale. There is no 
significant detection elsewhere. Whereas using the cross derivative 
coefficients the probability of detecting a non-gaussian signature at the 
third scale is 94.5\%. We apply the K-S test to the distribution of the excess
of kurtosis for the cross derivative and find a probability of 96.6\% of
detecting non-gaussianity. Since the K-S test compares the two distributions,
it is very sensitive to departures form gaussianity. It thus gives better
results on the detection of the non-gaussian signature.
\section{Discussion \& Conclusion}
The secondary anisotropies, due to CMB photon interactions, are superimposed on
the primary anisotropies which are directly related to the seeds of the 
cosmic structures. The primary anisotropies can 
be gaussian distributed (inflationary models) or can exhibit an intrinsic 
non-gaussian signature (topological defect models). In the context of future 
CMB observations (high sensitivity, high resolution and large sky coverage),
we will use the full information related to the CMB
temperature anisotropies, in particular the statistical information, to 
distinguish between the two main cosmological models. Similarly, studies
aiming at predicting and quantifying the foreground
contributions to the temperature anisotropies have to characterise the 
non-gaussian foreground signals in order to
subtract them before detailed CMB analysis. \par
In the present study we
investigate the tests of non-gaussianity when this is induced by 
secondary anisotropies, the primary anisotropies being gaussian distributed. 
We study the effects
arising from the interactions of the CMB photons and the ionised matter.
More specifically, we focus on two effects which dominate all the other
secondary effects of a scattering nature: the spatially
inhomogeneous re-ionisation which peaks at scales of a few tens of arcminutes 
to one degree and the SZ effect which dominates at the few arcminutes 
scale. In order to search for non-gaussianity we use discriminators based on 
the study of the statistical properties of the coefficients in a four level 
wavelet decomposition (Forni \& Aghanim 1999).
\par
The primary anisotropies are gaussian at all scales. Nonetheless, we
find a non-zero value of the multi-scale gradient excess of kurtosis, and 
hence first
derivatives, at the second decomposition scale which could be misinterpreted 
as a non-gaussian signature. This can be understood in the following way:
the window
function of the wavelet at this scale (centred around $l\sim2800$)
encompasses the cut off in the angular power 
spectrum. As a result the corresponding sample variance induces a non-zero
kurtosis for the multi-scale gradient coefficients. The presence of this
non-zero value depends on the cosmological
model as well as on the window filter that is the wavelet function. A similar 
non-zero value could exist at any decomposition scale where the CMB power 
spectrum has a sharp cut off. For the standard CDM model we use here, the
cut off occurs at the second scale. In the case where
the cosmological model has more power at small angular scales, or undergoes an
overall shift of the spectrum towards large multipoles, the sample variance 
effects decrease.
In the same way, we can use a wider wavelet which in turn decreases the 
sample effects. However, this attenuates the non-gaussian signature we search 
for. We apply a detection strategy proposed in Forni \& Aghanim (1999), which allows
the quantification of the detectability regardless of the power spectrum of 
the studied signal.\par
We have studied the case of secondary anisotropies induced by a spatially
inhomogeneous re-ionisation of the Universe. Assuming that this
was the only source of secondary anisotropies, we 
succeed in demonstrating its non-gaussian signature at the first and second
decomposition scales. 
However, inhomogeneous re-ionisation is far from being the only source of
anisotropies. The SZ effect due to galaxy clusters is known to be
the most common source which is related to the CMB photon scattering off free
electrons. In this study, we also take into account the
SZ effect of a predicted cluster population which we add to the 
primary CMB fluctuations and to the re-ionisation anisotropies. The 
non-gaussian foreground model is a worst case example because we do not remove 
any foregrounds. Owing to
its peculiar spectral signature the thermal effect is expected to be removed 
from the cosmological signal (temperature anisotropies). However, the
subtraction is not complete because almost 1/5 of the SZ effect contribution 
is due
to the kinetic SZ effect, which is spectrally indistinguishable from the
primary anisotropies, and there remains a significant non-gaussian foreground
contribution. 
In our study, we find that the dominant non-gaussian signal is due
to the SZ effect of clusters. The non-gaussian signature is found to be orders 
of magnitude larger than in the case without the SZ contribution and we 
clearly detect the non-gaussianity. The strong
non-gaussian signature, associated with the SZ effect, comes from the gas
profile of individual clusters. We have analysed temperature anisotropy maps
with different profiles (gaussian, $\beta$ profiles or even so point-like
sources) to which we add the primary gaussian anisotropies. As it is
very peaked at the centre, the cluster induces a sharp variation in the signal 
from the center to the
outskirts of the structure. In addition, an important fraction of the cluster
population is composed of unresolved point-like clusters. We thus find that
clusters represent the dominant non-gaussian foreground.
\par
We apply our statistical tests to Planck-like and MAP-like instrumental 
configurations in order to compare the capabilities of the two planned 
satellites
for detecting the non-gaussian signature induced by the secondary anisotropies
(mainly the SZ effect). For both configurations the fourth, and largest scale, 
shows no significant non-gaussianity due to the SZ 
contribution. In the MAP-like configuration the beam convolution
affects the first two decomposition scales. Therefore, 
we are only left with the third scale to search for non-gaussianity. At
the same time, the convolution rather sharply reduces the contributions at 
angular
scale associated with the third decomposition level. This induces a non-zero
excess of kurtosis. We apply our detection strategy to overcome the problem
and avoid a possible misinterpretation on non-gaussianity. We find no 
significant
detection of the non-gaussian signature at the third scale for the MAP-like 
configuration. By contrast, for the
Planck-like configuration, we detect the non-gaussian signature at the third
decomposition scale, the first and second ones being affected by the beam 
convolution.
\par\bigskip
We have shown that our statistical tests combined with a detection strategy
based on the characterisation of gaussian test maps, with same power spectrum 
as the non-gaussian studied process, are appropriate tools for 
demonstrating a non-gaussian signature. 
In a forthcoming paper, we will search for other discriminatory methods that 
allow two (or more) non-gaussian signals to be distinguished, in order to
subtract the non-gaussian signature of the secondary anisotropies
from the non-gaussian signature of the primary fluctuations. 
\begin{acknowledgements}
The authors would like to thank the referee A. Heavens for helpful 
comments that improved the paper
and P.G. Ferreira for kindly providing an IDL code
generating gaussian realisations, and for fruitful discussions. We also
wish to thank J.-L. Puget and F.R. Bouchet for helpful comments 
and A. Jones for his careful reading.
\end{acknowledgements}

%\bibliographystyle{/users1/mamd/tex/doctorat/natbib}

%\bibliography{/users1/nabila/articles/refer/all}
\end{document}